\documentclass[12pt]{article}
\normalbaselines
\title{Nonlinear Quantum Mechanics and Locality\thanks{Extended
    version of a talk presented at ``Symmetries in Science X'',
    Bregenz July 13-18, 1997.}}
\author{W.~L\"{u}cke$^1$ and P.~Nattermann$^2$\\ 
 $^1$Arnold Sommerfeld Institute for
 Mathematical Physics\\
 $^2$Institute for Theoretical Physics \\
 Technische Universit\"at Clausthal \\ D-38678
Clausthal, Federal Republic of Germany}
\date{July 30, 1997}
\def\Operator#1{\mathchoice
   {\mbox{\boldmath $#1$}}{\mbox{\boldmath $#1$}}
   {\mbox{\footnotesize \boldmath $#1$}}
   {\mbox{\footnotesize \boldmath $#1$}}} %kein \boldmath in \scriptsize
\font\boldit=cmbxti10 scaled \magstep1
\def\Notion#1{{\boldit #1\/}}
\def\rstop#1{\right.}
\def\bbbr{{\rm I\!R}} %reelle Zahlen
\def\aside#1{(#1)}
\def\stackscript#1{\mathop{\smash{#1}\vphantom{#1}}\limits}
\def\supp{\mbox{\rm supp\,}}
\def\jvV{\vec\jmath_{1,V}}
\def\jv{{\vec \jmath}}
\def\Nv{{\vec \nabla}}
\def\OA{{\Operator A}}
\def\Op#1{{\Operator #1}}
\def\xv{{\vec x}}
\def\norm#1{\left\| #1 \right\|}
\def\modulus#1{\left| #1 \right|}

\begin{document}

\maketitle
\begin{abstract}
It is shown that, in order to avoid unacceptable nonlocal effects, the
free parameters of the general {\sc Doebner-Goldin} equation have to be
chosen such that this nonlinear Schr\"odinger equation becomes Galilean
covariant.
\end{abstract}

\section{Introduction}

Usually linear equations in physics have the status of useful
approximations to actually nonlinear laws of nature. Therefore many
authors asked whether the fundamental linearity of quantum mechanics in
the form of the `superposition principle' plays a similar role. Also in
view of the persisting difficulties to combine the fundamental
principles of quantum mechanics with those of relativity into a rigorous
theory with nontrivial interaction it seems worthwhile to test nonlinear
modifications of ordinary quantum mechanics. For such reasons, many
authors suggested the addition of nonlinear terms to the linear
Schr\"odinger equation while maintaining the usual statistical
interpretation concerning the localization of the physical system (see,
e.g., \cite{MielnikNLQ,BiBiMy,HaBa,WeinTQ,DoGoGen}).

Unfortunately, the general interest in such theories was strongly
diminished by N.~{\sc Gisin}'s claim that every (`deterministic')
nonlinear Schr\"odinger equation leads to nonlocality of unacceptable
type \cite{Gisin2}, \cite{Gisin3}. However, his reasoning relies on the
tacit assumption -- not justified at all \cite{LueckeNL} -- that the
theory of measurements developed for the linear theory may be applied to
the nonlinear case, too. Therefore the question whether nonlinear
modifications of ordinary quantum mechanics may be physically
consistent deserves further investigation. Actually, the question is
whether, for a 2-particle system  with fixed initial conditions, the
nonlinearity allows to influence the position probability of particle 1
by acting on particle 2 if there is no explicit interaction between the
particles. In the following we will show that this possibility really
exists for some cases of the general {\sc Doebner-Goldin} equation, at
least.

This contribution is organized as follows. In Section \ref{SNLSE} we
will specify the type of nonlinear quantum mechanics we are going to
analyze. In Section \ref{SGisin} the central problem will be posed and
recent results by R.~{\sc Werner} related to this will be reported. In Section
\ref{SCheck}, finally, {\sc Werner}'s conjecture concerning nonlocality
of the general {\sc Doebner-Goldin} equation will be confirmed by simple
explicit calculations. We conclude with a short summary and further
perspectives.
\vskip 5mm

\section{Nonlinear Quantum Mechanics} \label{SNLSE}

Let us consider a typical nonlinear Schr\"odinger equation\footnote{We
use natural units, therefore $\hbar=1\,$.}
\begin{equation} \label{GNLS}
  \fbox{\fbox{$i\partial_t\Psi_t(\xv) = \Op H \Psi_t(\xv)
  +F_t(\Psi_t)(\xv)$}} 
\end{equation}
which is formally local in the sense that 
the nonlinearity $F_t$ added to the usual Schr\"odinger equation with
Hamiltonian $\Op H$ is some {\bf local} (non-linear) functional $F_t\,$,
$$
F_t[\Psi](\xv) = F_t[\Phi](\xv) \quad\forall\,\xv\notin\supp(\Psi-\Phi)\,,
$$
that should be `sufficiently small' in the sense that it does not
introduce too strong deviations from the predictions of the linear theory.
The question is whether the usual (nonrelativistic) quantum mechanical
interpretation
\begin{equation} \label{basic}
\fbox{\fbox{$\modulus{\Psi_t(\xv)}^2 = \left\{\begin{array}[c]{l}
\mbox{probability density for the system to be}\\
\mbox{localized around }\xv \mbox{ at time }t
\end{array}\rstop\}$}}
\end{equation}
may still be physically acceptable. Of course, (\ref{basic}) requires
the norm of solutions $\Psi_t$ of (\ref{GNLS}) to be $t$-independent.
This is automatically fulfilled if we restrict to nonlinearities of the
form
\begin{equation} \label{special}
F[\Psi] = R[\Psi]\,\Psi\;,\quad R[\Psi] = \overline{R[\Psi]}\,,
\end{equation}
since then (\ref{GNLS}) implies the ordinary continuity equation. In
view of the mentioned locality problem let us concentrate on  the case
of two noninteracting particles of different type in individual external
potentials $V_1,V_2:$
\begin{equation} \label{2P-NLS}
\begin{array}[c]{rcl}
\displaystyle i\partial_t \Psi^V_t(\xv_1,\xv_2)
&=&\displaystyle \left(
-\frac{1}{2m_1}\Delta_{\xv_1} + V_1(\xv_1,t)
-\frac{1}{2m_2}\Delta_{\xv_2} +
V_2(\xv_2,t)\right)\Psi^V_t(\xv_1,\xv_2)\\
&&\rule{0mm}{6mm}+ F\left(\Psi^V_t(\xv_1,\xv_2)\right)
\end{array}
\end{equation}
(here $\xv=(\xv_1,\xv_2$) and to nonlinearities of the {\sc
Bialynicki-Birula--Mycielski} type
\begin{equation} \label{BBM}
F_{\rm BB}(\Psi)=-\ln\rho\,\Psi
\end{equation}
or the {\sc Doebner-Goldin} type\footnote{The general {\sc
Doebner-Goldin} equation \cite{DoGo,DoGoGen} arises from this family by
nonlinear gauge transformations that do not change $\rho(\xv,t)$
\cite{NattD,NattT93}.}
\begin{equation} \label{DG}
F_{\rm DG}(\Psi)=\left(c_1 \frac{\Nv\cdot \vec J}{\rho} + c_2
\frac{\Delta\rho}{\rho} + c_3 \frac{\vec
J^2}{\rho^2}+ c_4 \frac{\vec J\cdot
\Nv\rho}{\rho^2} + c_5 \frac{(\Nv\rho)^2}{\rho^2}\right)\Psi\,,
\end{equation}
where we use the notation
$$
\rho = \modulus\Psi^2\;,\quad \vec J= \frac{1}{2i}\left(\overline{\Psi}\Nv\Psi
-\Psi\Nv\overline{\Psi}\right)\,,
$$
We assume that there are sufficiently many solutions of (\ref{2P-NLS})
for which the formal singularities, introduced especially by (\ref{DG}),
do not cause any problems (see \cite{CazenaveHaraux} and
\cite{Teismann}, in this connection). Interaction between the particles
and the case of identical particles will be discussed later.
\vskip 5mm

\section{The Locality Problem} \label{SGisin}

In both cases, (\ref{BBM}) and (\ref{DG}), $F$ is of the form
(\ref{special}) with
$$
R(\phi_1\otimes\phi_2) = R(\phi_1) + R(\phi_2)\,,
$$
and therefore
$$
\Psi^V_t(\xv_1,\xv_2) = \phi^{V_1}_t(\xv_1)\phi^{V_2}_t(\xv_2)
$$
is a solution of (\ref{2P-NLS}) whenever the $\phi^{V_j}_t$ are
solutions of the corresponding 1-particle equations
$$
i\partial_t \phi^{V_j}_t = \left(\Op H_j
+R\left(\phi^{V_j}_t\right)\right) \phi^{V_j}_t\,, 
$$
where
$$
\Op H_j \stackrel{\rm def}{=} -\frac{1}{2m_j}\Delta_\xv
+V_j(\xv,t)\,.
$$
This ensures that we cannot influence particle 1 by action on particle 2
by change of $V_2$ {\bf if} the fixed initial conditions are factorized.
However, most interesting features of quantum mechanics are connected
with entangled states (nonfactorized initial conditions). For
linear $F\,$, since the particles do not interact with each other, we
even have \Notion{full separability}:\footnote{Actually, one should
allow for magnetic fields.}
\begin{quote}
For arbitrarily fixed initial conditions,
the partial state of particle 1 does not depend on $V_2\,$.
\end{quote}
In other words:
\begin{quote}
$\left\langle \Psi_t \mid \OA\otimes\Op 1\mid \Psi_t \right\rangle$ does
not depend on $V_2$ for any self-adjoint operator in $L^2(\bbbr^3)\,$.
\end{quote}
That the latter statement is no longer true for nonlinear $F$ \aside{irrelevant Gisin
effect \cite{Gisin3}} does not mean that the former statement is wrong for  nonlinear
$F\,$, too \cite{LueckeNL}.
However, full separability should be equivalent to $V_2$-independence
of\footnote{Anyway, by (\ref{basic}), full separability implies this condition.}
\begin{equation} \label{rho}
\rho_{1,V}(\xv_1,t) \stackrel{\rm def}{=}\int
\modulus{\Psi^V_t(\xv_1,\xv_2)}^2{\rm d}\xv_2\,.
\end{equation}
If $\rho_{1,V}(\xv_1,t)$ changes with {\bf localized} (in space and time)
variations of $V_2$ then we have a \Notion{relevant Gisin effect}
\aside{unacceptable nonlocality}:
\begin{quote}
An arbitrarily small localized variation of $V_2$
may influence particle 1 at any distance by the same amount (just
translate $V_1$ and the initial condition w.r.t.\ $\xv_1$).
\end{quote}

Unfortunately, we do not have sufficient control on solutions of
(\ref{2P-NLS}). Therefore the only possibility to uncover relevant Gisin
effects, for the time being, is to determine $\left((\partial_t)^\nu
\rho_{1,V}\right)_{|_{t=0}}$ for fixed (entangled) initial conditions
and see whether this depends on $V_2$ for sufficiently large $\nu\,$.
Very recently Reinhard {\sc Werner} (Technical University Braunschweig)
performed a computer algebraic test of this sort for oscillator
potentials $V_j(\xv_j)=\kappa_j \norm{\xv_j}^2$ making the Ansatz
$$
\Psi^V_t(\xv_1,\xv_2) = \exp\left(-Q_t(\xv_1,\xv_2)\right)\,,
$$
where $Q_t$ is a time-dependent 2$^{\rm nd}$ order polynomial with
positive real part initial value $Q_0$ such that $\Psi^V_0(\xv_1,\xv_2)$
is not factorized.  {\sc Werner} found that $\left((\partial_t)^3
\rho_{1,V}\right)_{|_{t=0}}$ depends on $\kappa_2$ unless\footnote{Note
that condition (\ref{Werner}) is equivalent to Galilei covariance of
(\ref{2P-NLS}) for $F=F_{\rm DG}$ \cite{NattD} !}
\begin{equation} \label{Werner}
c_3=c_1+c_4=0
\end{equation}
Now, a variation of $\kappa_2$ means a nonlocalized variation of $V_2\,$.
But a global variation of $V_2$ may be approximated by a local variation.
Thus {\sc Werner} concluded that violation of (\ref{Werner}) implies
relevant Gisin effects.\footnote{Investigating $\left((\partial_t)^\nu
\rho_{1,V}\right)_{|_{t=0}}$ also for $\nu=4,\ldots,8$ {\sc Werner} did
not find anything more.}
\vskip 5mm

The only objection against the physical relevance of {\sc Werner}'s
result could be that in order to influence the position of particle 1
one might need local variations of $V_2$ of such strength that the
nonrelativistic equation (\ref{2P-NLS}), designed for sufficiently low
energies, is no longer applicable, anyway. Moreover, {\sc Werner}
himself admitted that Gaussian solutions might be too special and,
therefore, (\ref{Werner}) might not guarantee absence of relevant Gisin
effects. Therefore it is desirable to determine the $V_2$-dependent
part of $\left((\partial_t)^\nu \rho_{1,V}\right)_{|_{t=0}}$ for
essentially arbitrary initial conditions and potentials. This will be
done in the next Section for $\nu=3\,$, as a first step.
\vskip 5mm

\section{Confirmation of {\sc Werner}'s Results} \label{SCheck}

Obviously, as a consequence of the continuity equation
\begin{equation} \label{cont}
\partial_t \modulus{\Psi^V_t(\xv_1,\xv_2)}^2 + \Nv_{\xv_1}\cdot
\vec\jmath_{1,V}(\xv_1,\xv_2,t) + \Nv_{\xv_2}\cdot
\vec\jmath_{2.V}(\xv_1,\xv_2,t)=0\,,
\end{equation}
where
$$
\jv_{1,V}(\xv_1,\xv_2,t) = \Re\left(\overline{\Psi^V_t(\xv_1,\xv_2)}
\frac{1}{im_1}\Nv_{\xv_1} \Psi^V_t(\xv_1,\xv_2)\right)
$$
(and similarly for $\jv_{2,V}$) a relevant Gisin effect is equivalent to
nontrivial $V_2$-dependence of
\begin{equation} \label{rhodot}
\partial_t \rho_{1,V}(\xv_1,t) = -\int
\left(\Nv_{\xv_1}\cdot\jvV\right)(\xv_1,\xv_2,t)\,{\rm d}\xv_2\,.
\end{equation}
So the crucial question is whether
$$
\begin{array}[c]{rcl}
\partial_t \rho_{1,V}(\xv_1,t) \sim Q_V^\Phi(\xv_1,t) &\stackrel{\rm
def}{=}& \Im \int
\Nv_{\xv_1}\cdot\left(\overline{\Psi^V_t(\xv_1,\xv_2)}\Nv_{\xv_1}
\Psi^V_t(\xv_1,\xv_2)\right){\rm d}\xv_2\\
&=& \Im \int \overline{\Psi^V_t(\xv_1,\xv_2)}
\Delta_{\xv_1}\Psi^V_t(\xv_1,\xv_2)\,{\rm d}\xv_2
\end{array}
$$
is $V_2$-dependent for suitably fixed initial conditions $\Psi^V_0=\Phi\,$.
Like $Q_V^\Phi(\xv_1,0)\,$,
$$
\dot Q_V^\Phi(\xv_1,0) = \Re\int \left(\overline{(\Op
H_1\Phi+F(\Phi))}\Delta_{\xv_1} \Phi - \overline{\Phi} \Delta_{\xv_1}
\left(\Op H_1\Phi +F(\Phi)\right)\right){\rm d}\xv_2
$$
does not depend on $V_2$ and the only part of the r.h.s.\ of
$$
\begin{array}[c]{l}
-\ddot Q_V^\Phi(\xv_1,0)\\
= \Im \int \Bigl( \overline{\Op H_1(\Op H\Phi + F(\Phi))
+i\dot F(\Phi)}\Delta_{\xv_1}\Phi - \overline{\Op H_1\Phi+F(\Phi)}\Delta_{\xv_1}(\Op
H\Phi + F(\Phi))\\
\phantom{\Im \int \Bigl(x} - \overline{(\Op H\Phi +
F(\Phi))}\Delta_{\xv_1}(\Op H_1\Phi+F(\Phi)) + \overline{\Phi}
\Delta_{\xv_1}\left(\Op H_1(\Op H\Phi + F(\Phi))  +i\dot F(\Phi)\right)\Bigr){\rm
d}\xv_2
\end{array}
$$
\aside{$\Op H\stackrel{\rm def}{=}\Op H_1+\Op H_2$} which may depend on
the scalar potential $V_2$ is
$$
T_F^\Phi(\xv_1) \stackrel{\rm def}{=}
\Im \int \Bigl( \overline{i\dot F(\Phi)}\Delta_{\xv_1}\Phi + \overline{\Phi}
\Delta_{\xv_1} i\dot F(\Phi) -\overline{F(\Phi)}\Delta_{\xv_1}V_2\Phi -
\overline{V_2\Phi} \Delta_{\xv_1}F(\Phi)\Bigr){\rm d}\xv_2\,,
$$
where
$$
\dot F(\Phi)(\xv_1,\xv_2) \stackrel{\rm def}{=} \left(\partial_t
F(\Psi^V_t)(\xv_1,\xv_2)\right)_{|t=0} \,.
$$
By (\ref{special}) this gives
\begin{equation} \label{totest}
T_F^\Phi(\xv_1) = \Re\int\left(\overline{\Phi\dot
R}\Delta_1\Phi -\overline{\Phi}\Delta_1\left(\Phi\dot R\right)\right){\rm
d}\xv_2\,,
\end{equation}
where
$$
\dot R(\Phi)(\xv_1,\xv_2) \stackrel{\rm def}{=} \left(\partial_t
R(\Psi^V_t)(\xv_1,\xv_2)\right)_{|t=0} \,.
$$
For $R(\Psi) = G(\rho)\,$, therefore, (\ref{totest}) cannot depend on
$V_2$ since neither $\modulus{\Psi^V_0}^2$ nor
$\left(\partial_t\Psi^V_t\right)_{|t=0}$ does. In other words:
$$
\fbox{\parbox{13cm}{
For arbitrary initial conditions
$\left((\partial_t)^3 \rho_{1,V}\right)_{|_{t=0}}$ does not depend on
$V_2$ if the nonlinear functional $R$ is a real linear combination of
the three functionals
$$
R_{\rm BB}(\Psi) \stackrel{\rm def}{=} \ln\rho\;,\quad R_2(\Psi)
\stackrel{\rm def}{=} \frac{\Delta\rho}{\rho}\;, \quad R_5(\Psi)
\stackrel{\rm def}{=} \left(\frac{\Nv\rho}{\rho}\right)^2\,.
$$}}
$$
For
$$
R_4(\Psi) = \frac{\vec J\cdot\Nv\rho}{\rho^2}\,,
$$
however, we have
$$
\begin{array}[c]{l}
{\rm ess\,}\left(\partial_tR_4(\Psi^V_t)\right)_{|_{t=0}}\\
=\displaystyle \frac{\Nv\modulus\Phi^2}{\modulus\Phi^4}\cdot{\rm ess\,}\left(
\partial_t \frac{\overline{\Psi^V_t}\Nv \Psi^V_t
-\Psi^V_t\Nv\overline{\Psi^V_t}}{2i}\right)_{|_{t=0}}\\
=\displaystyle 2\frac{\Nv\modulus\Phi}{\modulus\Phi^3}\cdot\Re\left(
\overline{V_2\Phi}\Nv\Phi - \overline{\Phi}\Nv(V_2\Phi)\right)\\
=\displaystyle -2\frac{\Nv\modulus\Phi}{\modulus\Phi^3}\cdot\Re\left(
\overline{\Phi}[\Nv_2,V_2]_-\Phi\right)=
-2\frac{\Nv_2\modulus\Phi}{\modulus\Phi}
\cdot \Nv_2V_2\,,
\end{array}
$$
where `ess' means `$V_2$-dependent part of' and $\Phi=\Psi^V_0\,$.
Therefore the term
$$
{\rm ess\,}\left(\Phi\partial_tR_4(\Psi^V_t)\right)_{|_{t=0}} =
-2\frac{\Phi}{\modulus\Phi} \left(\Nv_2\modulus\Phi\right)
\cdot \Nv_2V_2
$$
to be inserted in the integral defining $T^\Phi_{F_4}$ is sufficiently
well behaved in order to take the limit of compactly supported $\Phi\,$.
Thus we may simplify our check by considering {\bf localized} $V_2$
fulfilling
\begin{equation} \label{specV}
r(\xv_1,\xv_1) \ne 0 \Longrightarrow V_2(\xv_2,0) = g x_2
\end{equation}
and initial conditions of the form
\begin{equation} \label{smoothinit}
\Phi(\xv_1,\xv_2) = e^{is(\xv_1,\xv_2)}r(\xv_1,\xv_2)\;,\quad
r,s\in C^\infty_0(\bbbr^3\times\bbbr^3,\bbbr)\,.
\end{equation}
Then
$$
{\rm ess}\left(\Psi^V_t\partial_tR_4(\Psi^V_t)\right)_{|_{t=0}} =
{\rm ess}\displaystyle\left(-2\frac{\Phi\Nv_2
\modulus\Phi}{\modulus\Phi}\cdot \Nv_2V_2\right) =
-2g\frac{\Phi}{\modulus\Phi}\partial_{x_2}\modulus{\Phi}
$$
and, consequently, the the essential part of $T^\Phi_{F_4}$ is
$$
\begin{array}[c]{rcl}
{\rm ess}\left(T^\Phi_{F_4}\right)&=& \displaystyle -2g\,\Re\int
\left(\frac{\overline{\Phi}\partial_{x_2}
\modulus\Phi}{\modulus\Phi}\Delta_1\Phi - \overline{\Phi}\Delta_1
\frac{\Phi\partial_{x_2}\modulus\Phi}{\modulus{\Phi}}\right) {\rm
d}\xv_2\\
&=& \displaystyle -2g\Re\int \left(
e^{-is}\left(\partial_{x_2}r\right)\Delta_{\xv_1}(e^{is}r) - e^{-is}r
\Delta_{\xv_1}\left(e^{is}\left(\partial_{x_2}r\right)\right)\right){\rm
d}\xv_2\,.
\end{array}
$$
To simplify things further, let us assume that\footnote{In fact,
(\ref{specs}) does not contribute to ${\rm
ess}\left(T^\Phi_{F_4}\right)\,$, but will be needed later.}
\begin{equation} \label{specs}
s(\xv_1\xv_2)=x_1x_2\,.
\end{equation}
Then
$$
\begin{array}[c]{rcl}
{\rm ess}\left(T^\Phi_{F_4}\right)&=&
\displaystyle -2g\Re\int \left(
e^{-is}\left(\partial_{x_2}r\right)\Delta_{\xv_1}(e^{is}r) - e^{-is}r
\Delta_{\xv_1}\left(e^{is}\left(\partial_{x_2}r\right)\right)\right){\rm
d}\xv_2\\
&=& \displaystyle -2g\int \left(
\left(\partial_{x_2}r\right)\left(-x_2^2r + \Delta_{\xv_1}r\right)
+r x_2^2 \partial_{x_2}r - r
\Delta_{\xv_1}\partial_{x_2}r\right)\xv_2\\
&=& \displaystyle -2g\int \left( (\partial_{x_2}r)\Delta_{\xv_1} r
-r\Delta_{\xv_1}\partial_{x_2}r\right){\rm d}\xv_2\\
&=& \displaystyle 4g\int r\Delta_{\xv_1}\partial_{x_2}r \,\,{\rm
d}\xv_2\\
&\ne& 0\quad\mbox{in general\footnotemark}\,
\end{array}
$$
\footnotetext{Note that
$$
4g\int r \Delta_1\partial_{x_2}r \,\,{\rm d}\xv_2 =
\mp 8 g\, \left(x_1r_1\partial_{x_1}r_1\right)\int r_2^2\,{\rm d}\xv_2\,.
$$
for
\begin{equation} \label{specr}
r(\xv_1,\xv_2) = (x_1\pm x_2)r_1(\xv_1)r_2(\xv_2)\,.
\end{equation}}
We may conclude:
$$
\fbox{\parbox{13cm}{If $\displaystyle F(\Psi) = \frac{\vec
J\cdot\Nv\rho}{\rho^2} \Psi$ then there are allowed initial conditions
$\Phi$ for which $\left((\partial_t)^3 \rho_{1,V}\right)_{|_{t=0}}$ may
be changed by localized variations of $V_2\,$.}}
$$
\vskip 5mm

\noindent
Instead of checking the case
$$
R(\Psi)= R_1(\Psi) \stackrel{\rm def}{=} \frac{\Nv\cdot \vec J}{\rho}
$$
it is more convenient to consider
$$
R(\Psi)= R_{1-4}(\Psi) \stackrel{\rm def}{=} \frac{1}{2i} \Delta
\ln\left(\Psi/\overline{\Psi}\right) = R_1(\Psi)-R_4(\Psi)
$$
where we have
$$
\begin{array}[c]{rcl}
{\rm ess\,}\left(\partial_tR_{1-4}(\Psi^V_t)\right)_{|_{t=0}}
&=&\displaystyle -\frac{1}{2} \Delta\,{\rm
ess}\left(\left(\overline{\Psi^V_t}/\Psi^V_t\right)
i\partial_t\left(\Psi^V_t/\overline{\Psi^V_t}\right)\right)_{|_{t=0}}\\
&=&\displaystyle -\frac{1}{2} \Delta\,{\rm ess}
\left(\left(i\partial_t\Psi^V_t\right)/\Psi^V_t+ 
\overline{\left(i\partial_t\Psi^V_t\right)}/\overline{\Psi^V_t}
\right)_{|_{t=0}}\\
&=& -\Delta_2 V_2
\end{array}
$$
and therefore
$$
\begin{array}[c]{rcl}
{\rm ess}\left(T^\Phi_{F_{1-4}}\right) &=& \displaystyle {\rm
ess}\left(\Re\int\left(\overline{\Phi\dot R_{1-4}} \Delta_1\Phi
-\overline{\Phi}\Delta_1\left(\Phi\dot R_{1-4}\right)\right){\rm
d}\xv_2\right)\\
&=& \displaystyle -{\rm
ess}\left(\Re\int\left(\left(\overline\Phi\Delta_2 V_2\right) \Delta_1\Phi
-\overline{\Phi}\Delta_1\left(\Phi\Delta_2 V_2\right)\right){\rm
d}\xv_2\right)\\
&=& 0\,,
\end{array}
$$
i.e.:
$$
\fbox{\parbox{13cm}{For arbitrary initial conditions,
$\left((\partial_t)^3 \rho_{1,V}\right)_{|_{t=0}}$ does {\bf not}
depend on $V_2$ if $R(\Psi)= R_1(\Psi)-R_4(\Psi)\,$.}}
$$
\vskip 5mm

\noindent
For 
$$
R_3(\Psi) = \left(\frac{\vec J}{\rho}\right)^2\,,
$$
finally, we have
$$
\begin{array}[c]{rcl}
{\rm ess\,}\left(\partial_tR_3(\Psi^V_t)\right)_{|_{t=0}}
&=& 2 \frac{\vec J}{\rho^2}\cdot {\rm ess}\left(\partial_t\vec J\right)\\
&=& -\frac{\vec J}{\rho^2} \left(\overline{\Phi}
[\Nv,V_2]_- \Phi + \Phi [\Nv,V_2]_-\overline\Phi\right)\\
&\stackscript{=}_{(\ref{specV})}& ig \, \frac{\overline\Phi\partial_2\Phi
-\Phi\partial_2\overline\Phi}{\rho}\,.
\end{array}
$$
Therefore the essential part of $T^\Phi_{F_3}$ is
$$
\begin{array}[c]{l}
\displaystyle -g\Im\int\left(\overline{\partial_2\Phi}\Delta_{\xv_1}\Phi
- \overline\Phi \frac{\partial_2\Phi}{\Phi}\Delta_{\xv_1}\Phi
+\overline\Phi\Delta_{\xv_1}\partial_2\Phi -\overline\Phi
\Delta_{\xv_1}\left(\Phi\overline{\left(\frac{\partial_2\Phi}{\Phi}\right)}
\right)\right){\rm d}\xv_2\\
=\displaystyle g\Im\int
\overline\Phi\left(\frac{\partial_2\Phi}{\Phi}\Delta_{\xv_1}\Phi
+\Delta_{\xv_1}\left(\Phi\overline{\partial_2\Phi/\Phi}\right)\right){\rm
 d}\xv_2\\
= g \Im
\int\left(e^{-2is}\partial_2\left(e^{is}r\right)\Delta_{\xv_1}(e^{is}r)
+ e^{-is}r\Delta_{\xv_1}\left(e^{+2is}\partial_2(e^{-is}r)\right)\right){\rm
d}\xv_2\\
\stackscript{=}_{(\ref{specs})} g \Im\int\Bigl(
e^{-is}(\partial_2r)\Delta_{\xv_1}(e^{is}r) + e^{-is}r
\Delta_{\xv_1}(e^{is}\partial_2r)\\
\phantom{\stackscript{=}_{(\ref{specs})} g \Im\int\Bigl(
e^{-is}(\partial_2r)\Delta_{\xv_1}(e^{is}r)} + \underbrace{ix_1re^{-is}
\Delta_{\xv_1}(e^{is}r) +
e^{-is}r\Delta_{\xv_1}(-ix_1e^{is}r)}_{\stackscript{=}_{(\ref{specV})}-2i
e^{-is}r\partial_1(e^{is}r) } \Bigr){\rm d}\xv_2\\
= g\int\left(2x_2(\partial_2r)(\partial_1r) + 2x_2 r\partial_1\partial_2
r -2r\partial_1 r\right){\rm d}\xv_2\,,
\end{array}
$$
i.e.:
\begin{equation} \label{TF3}
{\rm ess}\left(T^\Phi_{F_3}\right)= -4g\int r\partial_1 r\,{\rm d}\xv_2\,.
\end{equation}
Obviously, (\ref{TF3}) is functionally independent\footnote{For
instance, (\ref{TF3}) does not always vanish for factorized $r$ whereas
(\ref{TF4}) does.} of
\begin{equation} \label{TF4}
{\rm ess}\left(T^\Phi_{F_4}\right) = 4g\int
r\Delta_{\xv_1}\partial_{x_2}r \,\,{\rm d}\xv_2\,.
\end{equation}
Since, as shown in  \cite{NattD}, {\sc Werner}'s condition
(\ref{Werner}) is equivalent to Galilei invariance of the general {\sc
Doebner-Goldin} equation (equation (\ref{GNLS} with $F=F_{\rm DG}$) we
may conclude:
$$
\fbox{\fbox{\parbox{13cm}{For solutions $\Psi^V_t$ of (\ref{2P-NLS})
with
$$
F(\Psi)=\displaystyle\left(c_1 \frac{\Nv\! \cdot\! \vec J}{\rho} + c_2
\frac{\Delta\rho}{\rho} + c_3\! \left(\frac{\vec
J}{\rho}\right)^2 + c_4 \frac{\vec J \!\cdot\! \Nv\rho}{\rho^2} +
c_5 \left(\frac{\Nv\rho}{\rho}\right)^2\right)\Psi
$$
$\left(\partial_t^3\rho_{1,V}(\xv_1,t)\right)_{|_{t=0}}$
cannot be changed by local $V_2$-variations (for arbitrarily fixed initial
condition) if and only if the coefficients $c_\nu\in\bbbr$ are chosen
such that (\ref{2P-NLS}) is Galilei covariant.
}}}
$$
\vskip 5mm

\section{Summary} \label{SConc}

We have seen that the general {\sc Doebner-Goldin} equation has to be
Galilei invariant in order to avoid unacceptable nonlocalities for
noninteracting particles. Obviously, an interaction between the
particles that vanishes for infinite separation of the particles would
not have any influence on this conclusion. Similarly, since we
considered local variations of $V_2$ and since (\ref{specs}) and
(\ref{specr}) do not forbid any permutation symmetry of $\Phi\,$, the
same conclusion applies to pairs of identical particles.

Whether Galilei invariance protects the general {\sc Doebner-Goldin}
equation against relevant Gisin effects is not yet clarified. It may
well be that already
$\left(\partial_t^4\rho_{1,V}(\xv_1,t)\right)_{|_{t=0}}$ depends on
local $V_2$ variations even in the Galilei covariant case. The same, of
course, applies to the Bialynicki-Birula--Mycielski equation (equation
(\ref{GNLS}) with $F=F_{\rm BB}$).

Let us finally remark that even `full separability' would not yet be all
one would like to have:

For every 2-particle initial wave function $\Phi\,$, $\rho_{1,V}$ should
be the position probability density of a (possibly mixed) one-particle
state, i.e.\ there should exist a sequence of families of (unnormalized)
1-particle solutions $\psi^{V_1}_{\nu,t}$ with
$$
\sum_\nu \modulus{\psi^{V_1}_{\nu,t}(\xv_1)}^2 = \int
\modulus{\Psi^V_t(\xv_1,\xv_2)}^2 {\rm d}\xv_2\quad\forall\,V=(V_1,V_2)\,,
$$
where $\Psi^V_t$ denotes the corresponding family of 2-particle
solutions with $\Psi^V_0=\Phi\,$.

\end{document}